\documentclass{article}

\UseRawInputEncoding 

\newtheorem{theorem}{Theorem}[section]
\newtheorem{lemma}[theorem]{Lemma}
\newtheorem{corollary}[theorem]{Corollary}

\newtheorem{problem}[theorem]{Problem}

\newcommand\qed{\begin{flushright} {\bf q.e.d.} \end{flushright} }
\newcommand\prf{\noindent {\bf Proof :}}  

\newcommand\bits{\{0,1\}}
\newcommand\uu{{\bits^*}}

\newcommand\nnn{{\bits^{n+1}}}

\newcommand\ccc{{\bits^{c+1}}}

\newcommand\tru{{\mbox{{\bf tt}}}}
\newcommand\fw{{\Phi^w}}
\newcommand\flanw{{\varphi_{n,w}}}
\newcommand\transl{{\mbox{{\bf transl}}}}
\newcommand\fnw{f(1^{(n)}, w)}
\newcommand\pp{{\mbox{P}}}
\newcommand\np{{\mbox{NP}}}
\newcommand\conp{\mbox{coNP}}
\newcommand\ee{\mbox{E}}

\begin{document}
	
\title{A proof complexity conjecture and the Incompleteness theorem} 
	
	\author{Jan Kraj\'{\i}\v{c}ek}
	
	\date{Faculty of Mathematics and Physics\\
		Charles University\thanks{
			Sokolovsk\' a 83, Prague, 186 75,
			The Czech Republic,
			{\tt krajicek@karlin.mff.cuni.cz}\\
			ORCID: 0000-0003-0670-3957}}
	
	\maketitle

	\begin{abstract}
Given a sound first-order p-time theory $T$ capable of formalizing syntax of first-order logic we define a
p-time function $g_T$ that stretches all inputs by one bit and we use its properties to show that
$T$ must be incomplete. We leave it as an open problem whether for some $T$ the range of $g_T$
intersects all infinite $\np$ sets (i.e. whether it is a proof complexity generator hard for all proof systems).

A propositional version of the construction shows that at least one of the following three statements is true:
\begin{enumerate}
\item there is no p-optimal propositional proof system
(this is equivalent to the non-existence of a time-optimal 
propositional proof search algorithm),
	
\item $E \not\subseteq P/poly$,
	
\item there exists function $h$ that stretches all inputs by one bit, 
is computable in sub-exponential time 
and its range $Rng(h)$ intersects all infinite $\np$ sets.
\end{enumerate} 
	\end{abstract}
	
%

	\section{Introduction}

We investigate the old conjecture from the theory of proof complexity 
generators\footnote{We are not going to use anything from this theory but the interested reader may start with the introduction to \cite{Kra-strong} or with \cite[19.4]{prf}.}
that
says that there exists a generator hard for all proof systems. Its rudimentary version
can be stated without a reference to notions of the theory as follows:
\begin{itemize}
	\item {\em There exists a p-time function $g : \uu \rightarrow \uu$ stretching each input by one bit, $|g(u)| = |u| +1$, such that the range $Rng(g)$ of $g$ intersects all infinite
	$\np$-sets.}
\end{itemize}
We present a construction of a function $g_T$ (p-time and stretching)
based on provability in a first-order theory $T$ that is able to formalize syntax of first-order logic. Function $g_T$ has the property, assuming that $T$ is sound and complete, that
it intersects all infinite definable subsets of $\uu$. As that is clearly absurd (since $\uu \setminus Rng(g_T)$ is infinite and definable) this offers a proof of G\"{o}del's First Incompleteness theorem. 
We leave it as an open problem (Problem \ref{28.2.23b})
whether $g_T$ for some $T$ satisfies the conjecture above.

We then give a 
propositional version of the construction and use it to show that at least one of the following three statements has to be true:

\begin{enumerate}
	\item there is no p-optimal propositional proof system,
	
	\item $E \not\subseteq P/poly$,
	
	\item there exists function $h$ that stretches all inputs by one bit, 
	is computable in sub-exponential time $2^{O((\log n)^{\log \log n})}$
	and its range $Rng(h)$ intersects all infinite $\np$ sets.
\end{enumerate} 
We assume that the reader is familiar
with basic notions of logic and of computational and proof complexity
(all can be found in \cite{prf}).

\section{The construction}

We take as our basic theory $S^1_2$ of Buss \cite{Bus-book} (cf. \cite[9.3]{prf}), denoting its language simply $L$.
The language has a canonical interpretation in the standard model $\mathbf N$.
The theory is finitely axiomatizable and formalizes smoothly syntax of first-order 
logic. Language $L$ allows 
to define a natural syntactic hierarchy $\Sigma^b_i$ of bounded formulas that define
in $\mathbf N$ exactly corresponding levels $\Sigma^p_i$, for $i \geq 1$, 
of the polynomial time hierarchy. 

An $L$-formula $\Psi$ will be identified with the binary string naturally encoding it and 
$|\Psi|$ is the length of such a string. An $L$-theory $T$ is thus a subset of $\uu$, a set
of $L$-sentences, and it makes sense to say that it is p-time. It is well-known (and easy)
that each r.e. theory has a p-time axiomatization (a simple 
variant of Craig's trick, cf.\cite{Craig}).

If $u,v$ are two binary strings we denote by 
$u \subseteq_e v$ the fact that $u$ is an initial subword of $v$. The concatenation
of $u$ and $v$ will be denoted simply by $uv$. Both these relation and function are 
definable in $S^1_2$ by both $\Sigma^b_1$ and $\Pi^b_1$ formulas that are (provably in $S^1_2$)
equivalent. We shall assume that no formula is a proper prefix of another formula.

\bigskip

Let $T \supseteq S^1_2$ be a first-order theory in language $L$ 
that is sound (i.e. true in $\mathbf N$) and p-time. Define 
function $g_T$ as follows:

\begin{enumerate}

\item Given input $u$, $|u| = n$, find an $L$ formula $\Phi \subseteq_e u$ with one free variable $x$ such that $|\Phi| \le \log n$. (It is unique if it exists.)

\begin{itemize}
	\item If no such $\Phi$ exists, output $g_T(u) := \overline 0 \in \nnn$.
	 
	 \item Otherwise go to 2.

\end{itemize}

\item Put $c: = |\Phi|+1$. Going through all $w \in \ccc$ in lexicographic order, 
search for a $T$-proof of size $\le \log n$ of the following sentence $\fw$:
	\begin{equation} \label{28.2.23a}
	\exists y \forall x > y\ \Phi(x) \rightarrow \neg (w \subseteq_e x)\ .
	\end{equation}
	
\begin{itemize}
	\item If a proof is found for all $w$ output $g_T(u) := \overline 0 \in \nnn$.  
	
	\item Otherwise let 
	$w_0 \in \ccc$ be the first such $w$ such that no proof is found. Go to 3.

\end{itemize}	
	
\item Output  $g_T(u) := w_0 u_0 \in \nnn$, where $u = \Phi u_0$.	
\end{enumerate}

\begin{lemma} \label{28.2.23c}
	Function $g_T$ is p-time, stretches each input by one bit, and the complement of its range
	is infinite. 
\end{lemma}

The infinitude of the complement of the range follows as at most half of strings in
$\nnn$ are in the range.

\begin{theorem} \label{28.2.23d}
Let $A \subseteq \uu$ be an infinite $L$-definable set and assume that for some definition 
$\Phi$ of $A$ theory $T$ proves all true sentences $\fw$ as in (\ref{28.2.23a}), 
for $w \in \ccc$
where $c = |\Phi|$.
Then the range of function $g_T$ intersects $A$. 
\end{theorem}

\prf

Assume $A$ and $\Phi$ satisfy the hypothesis of the theorem. As $A$ is infinite some 
prefix $w$ has to appear infinitely many times as a prefix of words in $A$ and hence some sentence $\Phi^w$
is false. By the soundness of $T$ it cannot be provable in the theory.

Assuming that $T$ proves all true sentences $\fw$ let $\ell $ be a common upper bound to 
the size of some $T$-proofs of these true sentences. Then the algorithm computing $g_T(u)$
finds all of them if $n \geq 2^\ell$.

Putting this together, for $n \geq 2^\ell$ the algorithm finds always the same $w_0$ and this $w_0$ does indeed
show up infinitely many times in $A$. In particular, if $v \in \nnn \cap A$ is of the 
form $v = w_0 u_0$ and $n \geq 2^\ell$, then $v = g_T(\Phi u_0)$.

\qed

Applying the theorem to $A := \uu \setminus Rng(g)$ (and using Lemma \ref{28.2.23c}) yields 
the following version of G\"{o}del's First Incompleteness theorem.

\begin{corollary} 
	No sound, p-time $T \supseteq S^1_2$ is complete.
\end{corollary}

Note that the argument shows that for {\em each} formula $\Phi$ defining the complement, 
some true sentence $\fw$ as in (\ref{28.2.23a}) is unprovable in $T$. 
The complement of $Rng(g_T)$ is in $\conp$ and that leaves room for the following problem.

\begin{problem} \label{28.2.23b}
	For some $T$ as above, can each infinite $\np$ set be defined by {\em some} 
	$L$-formula $\Phi$
	such that all true sentences $\fw$ as in (\ref{28.2.23a}) are provable in $T$?  
\end{problem}
The affirmative answer would imply by Theorem \ref{28.2.23d}
that $Rng(g_T)$ intersects all infinite $\np$ sets and hence $g_T$
solves the proof complexity
conjecture mentioned at the beginning of the paper, 
and thus $\np \neq \conp$. Note that, for each $T$, it is easy to define
even as simple set as 
$$
\{1 u\ |\ u \in \uu\}
$$ 
by a formula $\Phi$ such that $T$ does not prove that no string in it starts with $0$.
But in the problem we do not ask if there is {\em one definition} leading to unprovability but
whether {\em all definitions} of the set lead to it. 

\section{Down to propositional logic} \label{11.3.23a}

The reason why the algorithm computing $g_T$ searches for $T$-proofs of formulas $\fw$
rather than of $\neg \fw$ which may seem more natural is that $\np$ sets can be defined by $\Sigma^b_1$-formulas $\Phi$ and for those, after substituting a witness for $y$,
$\fw$ becomes a $\Pi^b_1$-formula. Hence one can apply propositional translation
(cf. \cite{Coo75} or \cite[12.3]{prf}) and 
hope to take the whole argument down to propositional logic,
replacing the incompleteness by lengths-of-proofs lower bounds. 
There are technical complications along this ideal route but we are at least able to
combine the general idea with a construction akin 
to that underlying \cite[Thm.2.1]{Kra-di}\footnote{That theorem is similar in form 
to Theorem \ref{11.3.23b} but with 2) replaced by $\ee \not\subseteq Size(2^{o(n_)})$
and 3) replaced by $\np \neq \conp$.}

and to prove the following statement.

\begin{theorem} \label{11.3.23b}
	At least one of the following three statements is true:
\begin{enumerate}
	\item there is no p-optimal propositional proof system,
	
	\item $E \not\subseteq P/poly$,
	
	\item there exists function $h$ that stretches all inputs by one bit, 
	is computable in sub-exponential time $2^{O((\log n)^{\log \log n})}$
	and its range $Rng(h)$ intersects all infinite $\np$ sets.
\end{enumerate}

\end{theorem}
Note the first statement is
by \cite[Thm.2.4]{Kra-prfsearch} equivalent to the non-existence of 
a time-optimal propositional proof search algorithm.

\bigskip

Before starting the proof we need to recall a fact about propositional translations of
$\Pi^b_1$-formulas. For $\Phi(x) \in \Sigma^b_1$, $c :=|\Phi|$ and $w \in \ccc$, 
and $n \geq 1$ let $\flanw$ be the canonical propositional formula expressing that
$$
(|x| = n+1 \wedge \Phi(x)) \rightarrow \neg w \subseteq_{e} x\ .
$$ 
We use the qualification {\em canonical} because the formula can be defined using
the canonical propositional translation $||\dots||^{n+1}$ 
(cf. \cite[12.3]{prf} or \cite{Coo75}) applied to $\fw$ after instantiating first
$y$ by $1^{(n)}$. Formula $\flanw$ has $n+1$ atoms for bits of $x$ and $n^{O(1)}$ atoms
encoding a potential witness to $\Phi(x)$ together with the p-time computation that 
it is correct. For any fixed $\Phi$ the size of $\flanw$ (with $w \in \ccc$)
is polynomial in $n$ and, in fact, the formulas are very uniform (cf. \cite[[19.1]{prf}).
We shall need only the following fact.

\begin{lemma} \label{11.3.23c}
	There is an algorithm $\transl$ that upon receiving as inputs
	a $\Sigma^b_1$-formula $\Phi$, $w \in \ccc $ where $c := |\Phi|$ and 
	$1^{(n)}$, $n \geq 1$, outputs $\flanw$ such that 
	$$
	(|x| = n+1 \wedge \Phi(x)) \rightarrow \neg w \subseteq_{e} x\ .
	$$ 
	is universally valid iff $\flanw$ is a tautology. In addition, for any fixed $\Phi$
	the algorithm runs in time polynomial in $n$, for $n > |\Phi|$. 
\end{lemma}

\noindent 
{\bf Proof of Theorem \ref{11.3.23b}:}

To prove the theorem we shall assume that statements 
1) and 2) fail and (using that assumption) we construct
function $h$ satisfying statement 3). Our strategy is akin in part to that
of the proof of \cite[Thm.2.1]{Kra-di}.

For a fixed $\Phi$ assume that formulas $\flanw$ are valid for $n \geq n_0$. By 
Lemma \ref{11.3.23c} they are computed by $\transl(\Phi, w, 1^{(n)})$ in p-time. 
Hence we can consider
the pair $1^{(n)}, w$ to be a proof (in an ad hoc defined proof system)
of $\flanw$ for $n \geq n_0$

Assuming that statement 1) fails and $P$ is a p-optimal proof system
we get a p-time function $f$ that from $1^{(n)}, w$, $n \geq n_0$, computes a 
$P$-proof $f(1^{(n)}, w)$ of $\flanw$. Let $|\fnw| \le n^\ell$ where $\ell $ is a constant
(depending on $\Phi$). The function that from $n,w,i$, with $i \le n^\ell$, computes
the $i$-th bit of $\fnw$ is in the computational class $\ee$.

We would like to check the validity of $\flanw$ by checking the $P$-proof $\fnw$
but we (i.e. the algorithm that will compute $h$) cannot construct $f$ from $\Phi$.
Here the assumption that statement 2) fails too, i.e. that
$E \subseteq \pp/poly$,  will help us. 
By this assumption $\fnw$ is the truth-table $\tru(D)$ (i.e. the lexico-graphically 
ordered list of values of circuit $D$ on all inputs) of some circuit with
$\log n + c + \ell \log n \le (2+\ell)\log n$ inputs and of size
$|D| \le (\log n)^{O(\ell)}$. In particular, for all $\ell$ (i.e. for all 
$\Phi \in \Sigma^b_1$) we have\footnote{Note that the function $\log \log n$ 
bounding $\ell$ can be replaced by any $\omega(1)$ time-constructible function, 
making the time needed to compute function $h$ closer to quasi-polynomial.}
$|D| \le (\log n)^{\log \log n}$ for $n >> 1$.
Hence it is enough to look for $P$-proofs among $\tru(D)$ for circuits of 
at most this size.

We can now define function $h_P$ in a way analogous to the definition of function 
$g_T$. Namely:

\begin{enumerate}
	
	\item Given input $u$, $|u| = n$, find a $\Sigma^b_1$-formula $\Phi \subseteq_e u$ with one free variable $x$ such that $|\Phi| \le \log n$. (It is unique if it exists.)
	
	\begin{itemize}
		\item If no such $\Phi$ exists, output $h_P(u) := \overline 0 \in \nnn$.
		
		\item Otherwise go to 2.

	\end{itemize}
	
	\item Put $c: = |\Phi|+1$. Going through all $w \in \ccc$ in lexicographic order, 
	do the following.
	
	Using $\transl$ compute formula $\flanw$. If the computation does not halt
		in time $\le n^{\log n}$ stop and output $h_P(u) = \overline 0 \in \nnn$.
	Otherwise search for a $P$-proof of formula $\flanw$ by going systematically through
	all circuits $D$ with $\le \log n \cdot \log \log n$ inputs and of size 
		$\le (\log n)^{\log \log n}$
		until some $\tru(D)$ is a P-proof of $\flanw$.

	\begin{itemize}
		\item If a proof is found for all $w \in \ccc$ output 
		$h_P(u) := \overline 0 \in \nnn$.  
		
		\item Otherwise let 
		$w_0 \in \ccc$ be the first such $w$ such that no $P$-proof is found. Go to 3.

	\end{itemize}	
	
	\item Output  $h_P(u) := w_0 u_0 \in \nnn$, where $u = \Phi u_0$.	
\end{enumerate}
It is clear from the construction that
function $h_P$ stretches each input by one bit (and hence the complement of its range
is infinite)  and that 
$$
Rng(h_P) \cap \{x \in \nnn\ |\ \Phi(x)\} \neq \emptyset
$$
for any $\Phi(x) \in \Sigma^b_1$ and $n >> 1$.

The time needed for the computation of $h_P(u)$ is $O(n)$ for step 1 and for 
step 2 it is bounded above by
$$
2^{c+1} \cdot n^{\log n} \cdot 2^{(\log n)^{\log \log n}} \cdot 
2^{O((\log n)^{\log \log n})} 
\le 2^{O((\log n)^{\log \log n})} \ .
$$
The first factor bounds the number of $w$, the second bounds the time needed to 
compute $\flanw$, the third bounds the number of circuits $D$ and the fourth one
bounds the time needed to compute $\tru(D)$ and to check wether it is a $P$-proof
of $\flanw$ (this is p-time in $|\tru(D)|$).

\qed

\bigskip

\noindent
{\large {\bf Acknowledgments:}} 
Section \ref{11.3.23a} owns its existence to J.Pich (Oxford) who suggested I include
some propositional version of the construction.

\end{document}